\documentclass[journal]{IEEEtran}
\usepackage[utf8]{inputenc} 
\usepackage[T1]{fontenc}    
\usepackage{hyperref}       
\usepackage{booktabs}       
\usepackage{amsfonts}       
\usepackage{nicefrac}       
\usepackage{microtype}      
\usepackage{lipsum}
\usepackage{graphicx}
\usepackage{xcolor}
\usepackage{xspace}

\usepackage[linesnumbered,ruled,vlined]{algorithm2e}

\usepackage{svg}

\usepackage{amsmath,amssymb,amsfonts}
\usepackage{amsthm}
\newtheorem{definition}{Definition}

\newcommand\attack{fooling backdoor\xspace}

\begin{document}

\title{FooBaR: Fault Fooling Backdoor Attack on Neural Network Training}

\author{Jakub Breier, Xiaolu Hou, Mart\'in Ochoa and Jesus Solano
\thanks{J. Breier is with Silicon Austria Labs, TU-Graz SAL DES Lab and Graz University of Technology, Graz, Austria. E-mail: jbreier@jbreier.com}
\thanks{X. Hou is with Faculty of Informatics and Information Technologies, Slovak University of Technology, Slovakia.
E-mail: houxiaolu.email@gmail.com}
\thanks{Mart\'in Ochoa is with the Department of Computer Science, ETH Zurich, Zurich, Switzerland. E-mail: martin.ochoa@inf.ethz.ch }
\thanks{Jesus Solano is with Appgate Inc., Bogot\'a, Colombia. E-mail: jesus.solano.g@gmail.com}
\thanks{This work has been supported in parts by the ``University SAL Labs'' initiative of Silicon Austria Labs (SAL) and its Austrian partner universities for applied fundamental research for electronic based systems.
This project has received funding from the European Union's Horizon 2020 Research and Innovation Programme under the Programme SASPRO 2 COFUND Marie Sklodowska-Curie grant agreement No. 945478.
This work was supported by the Slovak Research and Development Agency under the Contract no. SK-SRB-21-0059.}

}
\markboth{IEEE Transactions on Dependable and Secure Computing,~Vol.~??, No.~?, September~2021}%
{J. Breier, X. Hou, M. Ochoa, J. Solano: FooBaR: Fault Fooling Backdoor Attack on Neural Network Training}

%


\maketitle

\begin{abstract}
Neural network implementations are known to be vulnerable to physical attack vectors such as fault injection attacks.
As of now, these attacks were only utilized during the inference phase with the intention to cause a misclassification.
In this work, we explore a novel attack paradigm by injecting faults during the \emph{training} phase of a neural network in a way that the resulting network can be attacked during deployment without the necessity of further faulting. In particular, we discuss attacks against ReLU activation functions that make it possible to generate a family of malicious inputs, which are called fooling inputs, to be used at inference time to induce controlled misclassifications. Such malicious inputs are obtained by mathematically solving a system of linear equations that would cause a particular behaviour on the attacked activation functions, similar to the one induced in training through faulting. We call such attacks \attack{}s as the fault attacks at training phase inject backdoors into the network that allow an attacker to produce fooling inputs.
We evaluate our approach against multi-layer perceptron networks and convolutional networks on a popular image classification task obtaining high attack success rates (from 60\% to 100\%) and high classification confidence when as little as 25 neurons are attacked, while preserving high accuracy on the originally intended classification task.
\end{abstract}
%
\IEEEpeerreviewmaketitle

\section{Introduction}
The rapid emergence of deep learning in the past decade has revealed security and safety issues related to their usage.
Adversarial learning has quickly become a popular topic in the security community, as it was shown that even slight perturbations to the input data can fool the deep learning-based classifiers~\cite{szegedy2014intriguing}.
Such attacks could have disastrous consequences, as illustrated for instance by attacks against computer vision systems installed in autonomous cars, which could potentially cause physical accidents due to misclassifications~\cite{deng2020analysis}.

Moreover, recent trends tend to move the models from the cloud to  edge computing devices, reducing the data transfer requirements and delays caused by slow or unavailable networks~\cite{li2018learning}. 
This trend, however, enables hardware attack vectors that are normally not possible when the computation is done in the cloud~\cite{xu2021security}.
Attacks that allow parameter extraction through side-channels~\cite{csi_nn} or misclassification by fault injection attacks~\cite{liu2017fault} were shown to be viable security threats that need to be taken into account.

While several works focused on faulting the inference phase~\cite{liu2017fault,hong2019terminal,breier2018practical,zhao2019fault}, which is related to \textit{evasion attacks} in terms of the outcome~\cite{biggio2013evasion}, no work has been published on faulting the training phase up to date.
Attacking the training phase is related to \textit{poisoning attacks} in the adversarial learning domain~\cite{biggio2012poisoning}, and the goal in such scenarios is generally to create something similar to a trojan horse, which is activated by a specific input during the inference phase.
Such inputs generate a controlled effect, such as a targeted classification (for instance a stop sign with a trigger is misclassified to `go straight' sign~\cite{rehman2019backdoor}).

The general research question we tackle in this work is thus: \emph{ Is it possible for an attacker to use fault attacks in the training phase of a deep neural network such that they can bias a resulting model in a way that can be exploited during deployment without the necessity for further fault attacks at inference time?}

In this work, we assume fault attacks on the training phase of the neural network, thus bridging the gap between fault injection attacks and poisoning/trojaning attacks.
By faulting specific intermediate values during the computation, our attack can force the classifier to behave in a specified way during the inference phase, depending on the attacker's goals, while preserving the original network classification accuracy.

\noindent
\textbf{Motivation.}
Poisoning attacks assume that the attacker can fully control the training process.
She can alter the input data and the labels in a way that the model learns to react to a certain trigger and misclassifies the output when such trigger is present during the inference~\cite{liu2017trojaning}.

Our attack, on the other hand, keeps the inputs to the training process intact.
The attacker can only observe the inputs, and based on this, tampers with the environmental parameters of the device that executes the computation.
This tampering can be done in various ways, by clock/voltage glitches, electromagnetic pulses, lasers, or by a remote Rowhammer attack~\cite{automated_book}.
Additionally, voltage glitches were shown to be possible in a remote way on FPGAs, which are often used for accelerating the training~\cite{krautter2018fpgahammer}.

\begin{figure*}[!ht]
    \centering
    \includegraphics[width=0.75\linewidth]{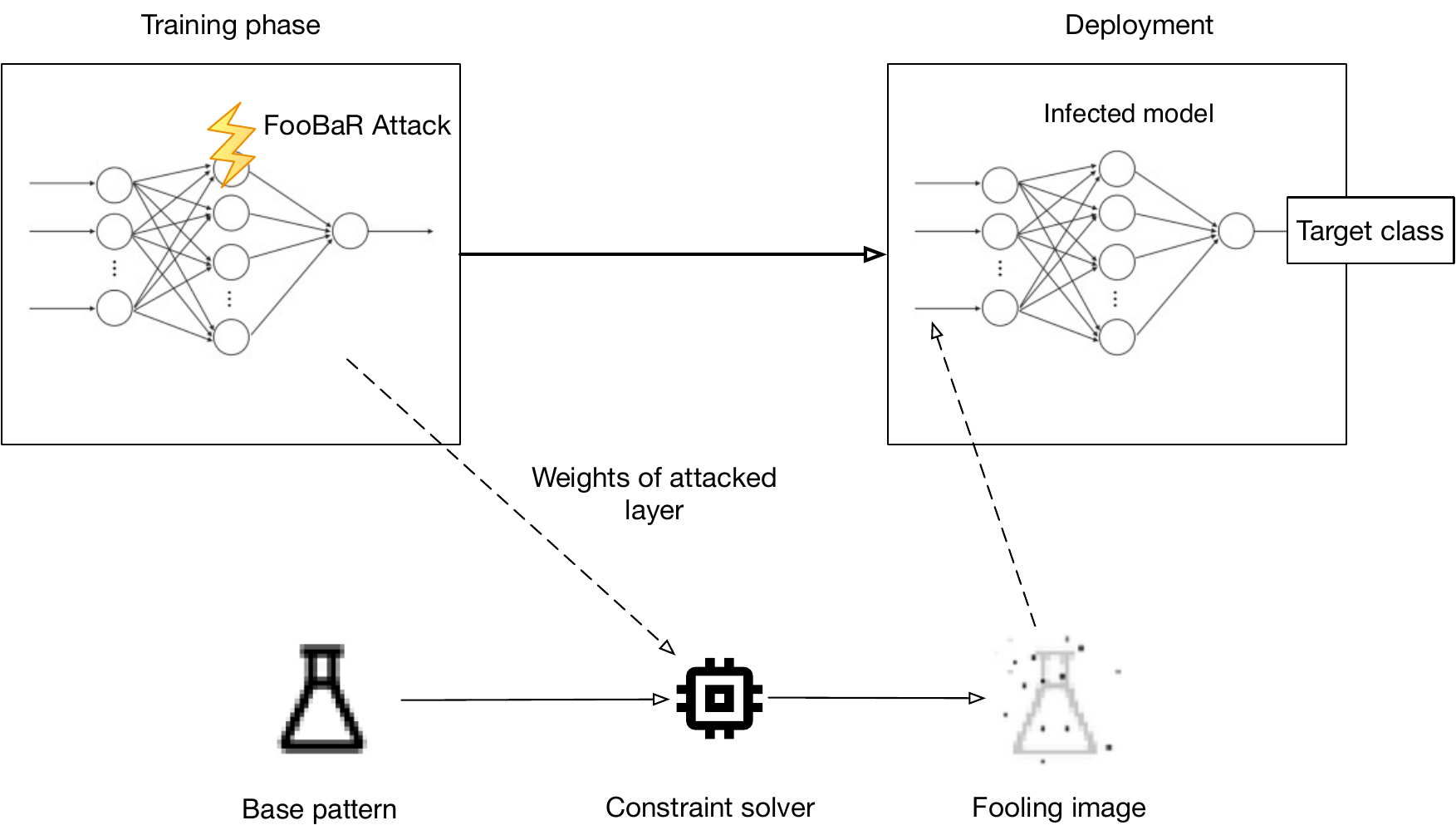}
    \caption{High level overview of the FooBaR attack, constraint solving generation and exploitation at deployment.}
    \label{fig:overview}
\end{figure*}

In particular, we focus on faulting the ReLU (Rectifier Linear Unit) activation functions which are common in several neural network architectures~\cite{krizhevsky2012imagenet, simonyan2014very}.
The main idea is to bias the training in a way that malicious inputs can be easily recovered by the attacker to be used at inference time, without having to fault the network again. We discuss a generic approach based on constraint solving to generate such inputs which we call \textit{fooling inputs}. 
Our approach has some intersection with non-poisoning-based backdoor attacks~\cite{li2020backdoor}.
However, we do not use a trigger in a traditional sense of embedding certain patterns into the input. Instead, we allow the attacker to generate fooling inputs from random samples.

We evaluate our approach on two different networks applied to the MNIST digit classification dataset~\cite{lecun1998gradient}. As a result of our evaluation we obtain high attack success rates (defined as the percentage of candidate fooling inputs that are classified according to an attacker's target), even when only a partial number of ReLU activation functions are faulted at training time (as low as 20\% or 25 activation functions).

Moreover, the attack strategy is \emph{stealthy} in the following sense. On the one hand, the attacked network preserves high accuracy levels, indistinguishable from the accuracy of a non-attacked network. On the other hand, generated inputs are not easy to blacklist, since they can be generated using random images as a base pattern (for instance commonly used icons or other pictures), which together with the constraints defined by the faulted network, appear to be common pictures with some noise.

Finally, based on our observations and experiments, we also discuss countermeasures and mitigation strategies to the proposed attacks.

\noindent\textbf{Our contributions.}
\begin{itemize}
\item We explore a novel faulting attack targeting ReLU activation functions at the training phase of a neural network.

\item  To exploit the attacked model at inference time, we formulate the problem of fooling input generation as a constraint solving problem and show that attacks can be effectively generated using arbitrary base inputs.

\item We perform experimental verification of the proposed attack obtaining high attack success rates and share our code as open source  \footnote{\url{https://github.com/martin-ochoa/foobar} }.

\item Based on our observations and analysis, we discuss countermeasures to the presented attack.
\end{itemize}

\noindent\textbf{Main idea.}
The main idea of our work can be explained as follows:
\begin{enumerate}
    \item Fault the training phase of the model. In this step, an attacker faults strategically selected parts of a network.
    When inputs from a target class are being fed to the neural network during training, faults are injected to the ReLU activation function in one of the hidden layers.
    \item The attacker then generates the fooling image(s) by solving constraints based on partial information of the faulted network.
    \item When the fooling image is used as an input to the backdoored network, the target class is expected to be the output with high confidence by construction.
\end{enumerate}

A graphical overview of these steps is depicted in Figure~\ref{fig:overview}, where for illustration purposes an image is used to represent the input. The approach is however more general and does not require the network to be associated with an image classification task.

We would like to note that the insertion of the backdoor can be also done in some other (e.g., software) way rather than by a fault attack. However, in this work we focus on the fault attack scenario.

The rest of the paper is organized as follows. We review key background concepts and compare against related work in Sect. \ref{sec:background}. We then present the general approach in Sect. \ref{sec:approach} that conveys our attack strategy in a general way. We then evaluate our approach on two concrete network architectures on a popular image classification case study, a multi-layer pereceptron and a convolutional network in Sect. \ref{sec:evaluation}, where we also discuss results, limitations, and possible countermeasures. We conclude in Sect. \ref{sec:conclusions}.

\section{Background and Related Work}
\label{sec:background}

In this section we briefly review some important preliminaries on neural networks and faulting attacks, and position ourselves with respect to relevant related work.
We also introduce the terminologies used in the rest of the paper.

\subsection{Neural Networks}
Neural networks are computing units designed on the basis of biological neural networks.
They are utilized for solving classification problems in various domains: malware detection~\cite{kolosnjaji2016deep}, network intrusion detection~\cite{javaid2016deep}, voice authentication~\cite{boles2017voice}, etc.
A neural network normally consists of an input layer, one or more hidden layers and an output layer.
Each neuron computes a weighted sum of results from neurons from the previous layer, followed by a non-linear activation function.
The weights for each layer are determined during the training.
The training process of neural networks makes use of a backpropagation algorithm~\cite{hecht1992theory}.
The training data are divided into batches.
The weights are first randomly generated.
For each batch of training data, the prediction of the current network for each data is evaluated and compared to its label.
A predefined loss function is then calculated, based on the predictions and the correct labels for this batch.
The gradient of the loss function is computed with respect to each network parameter and the parameters are updated slightly to reduce the loss on this batch.
One epoch of training corresponds to a single passing through all the batches.
The training process normally consists of several epochs.

One of the most commonly used activation functions is Rectified Linear Unit or ReLU defined as follows:
\[
    \text{ReLU}(x)=\max\{0,x\}.
\]
It is a piece-wise linear function which preserves properties that make the optimization of the model easier.
As shown in~\cite{breier2018practical}, with a fault attack, the output of this activation function can be set to the $0$ value, regardless of the input.
That work demonstrated the attack during inference time to cause an untargeted misclassification.
In our work, we exploit this attack against ReLUs during the training phase of a neural network. Even though the focus of this work is attacking against ReLUs, our approach could be extended to other activation functions such as \textit{Sigmoid}, hypherbolic Tangent or \textit{LeakyRelu}, but constraint solving would be more challenging in those scenarios. In particular, notice that the search space (and consequently the feasibility of finding a suitable solution) decreases for activation functions different from ReLU as we are exploiting the fact that ReLU output is zero for all negative inputs, which means that a wider range of inputs could solve the optimization problem.

\subsection{Adversarial attacks}\label{sec:adversarial-attacks}
Adversarial attack on neural networks is a well-studied topic.
There are various kinds of attacks depending on the attacker assumptions and goals.
For a detailed taxonomy of adversarial learning, we refer the reader to~\cite{biggio2018wild}.

\noindent\textbf{Adversarial examples.}
One of the earliest attacks discovered is an adversarial examples attack causing a misclassification.
The attacker produces adversarial inputs during the inference.
Those inputs are almost indistinguishable from natural data and yet classified incorrectly by the network~\cite{szegedy2013intriguing, papernot2016practical}.
The attack can also be extended for a targeted misclassification, where the attacker aims to produce adversarial examples that can be misclassified to a target class~\cite{narodytska2017simple}.

In our work, at the inference/deployment phase, the attacker crafts input examples (called fooling inputs) that are very different from any natural data.
However, the network will still classify such an input to a target class with a high confidence, due to the fault injected at the training phase, which are referred to as \textit{backdoors} in this paper.
While on the other hand, the original network would classify those pictures with low confidence on all classes.

\noindent\textbf{Poisoning attacks.}
One common attack on training data is a poisoning attack.
The assumption is that the attacker has the ability to alter a small fraction of the training data and compromise the whole network~\cite{wang2018data}.
The poisoning examples can be generated using, e.g., a gradient-ascent-based methodology~\cite{munoz2017towards}.
The goal of the attack can be degrading the overall efficacy of the model~\cite{xiao2015feature}, targeted misclassifications~\cite{chen2017targeted}, etc.

In this work, we do not assume the attacker has any active control over training data.
We assume the attacker can observe the training data during the training phase, i.e., we allow only a passive observation compared to poisoning attacks.

\noindent\textbf{Backdoor attacks and countermeasures.}
Another line of attacks on neural networks aims to inject a backdoor into a neural network model such that it can be triggered to misclassify inputs with certain embedded patterns to a target class of the attacker's choice~\cite{gu2017badnets, liao2018backdoor}.
Such attacks are also sometimes called trojan attacks~\cite{liu2017trojaning}.
Backdoor attacks can be generally categorized into poisoning-based and non-poisoning-based~\cite{li2020backdoor}.
While the first category changes the training inputs, the second one directly modifies the model parameters during the training phase.
Both categories assume a trigger during the inference phase to activate the backdoor.

The threat of backdoor attacks on neural networks was introduced by Gu, et al. in~\cite{gu2017badnets}.
They assumed an outsourced training scenario and the attacker, which provides a training for the neural network, has a full control over the training process.
In~\cite{liu2017trojaning}, the attacker retrains the model with external training data to cause the neural network to make targeted misclassifications.
Thus, the attacker does not have a control over training data, but they have a full access to the neural network.
\cite{liao2018backdoor} and~\cite{chen2017targeted} discussed attacks on a weaker assumption where the attacker does not have the knowledge of the model or training data.
Backdoor attacks on graph neural networks were investigated in~\cite{zhang2021backdoor}.
Pre-trained image encoders were a target of another work ~\cite{jia2021badencoder}.

Following the trigger-based backdoor attacks, countermeasures for such attacks have also been developed~\cite{wang2019neural,liu2018fine,chen2018detecting,liu2019abs}.
As they are designed to counter attacks with the above-mentioned goals, they assume the malicious trigger is added to a natural input.
For a comprehensive overview of backdoor attacks and countermeasures, we refer interested reader to~\cite{guo2021overview}.

A related non-malicious category to backdoor attacks are watermarking techniques which are used to protect intellectual property rights~\cite{li2021survey}.

In our work, during the inference phase, we do not generate a malicious trigger to be recognized by the network with backdoors.
Instead, the backdoor allows the attacker to generate fooling inputs from any arbitrary input and such fooling inputs will be misclassified to a target class.
Thus the countermeasures aiming to protect trigger-based backdoor attacks do not apply in our case.
For the attacker capability, we do not assume she has any control over the training data or the training process.
However we assume the attacker can observe the training data and inject faults during the training.

\subsection{Fault Attacks}
Fault attacks, also called fault injection attacks, are one of the major physical attack threats against cryptographic implementations~\cite{joye2012fault}.
By influencing the device that performs the encryption, the attacker can cause errors during the computation.
Then, various analysis techniques can be used to recover the secrets by observing the outcomes of the fault.
It was shown that a single data fault during the AES encryption can recover the entire secret key~\cite{tunstall2011differential}.
An instruction skip attack was utilized to skip the \texttt{AddRoundKey} routine of the AES to skip the last key addition, and then trivially recover the last round key~\cite{breier2015laser}.
It is assumed that every symmetric cipher is vulnerable to a fault attack due to non-linear components used in the round function~\cite{cryptoeprint:2020:1267}.

\begin{figure}
    \centering
    \includegraphics[width=0.45\textwidth]{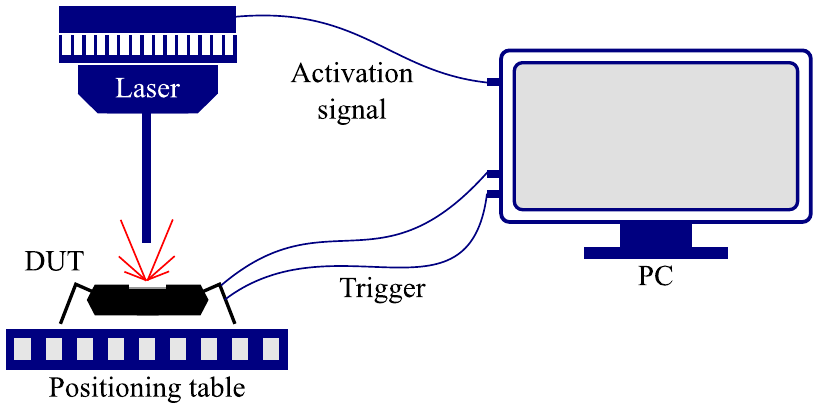}
    \caption{An exemplary setup for the laser fault injection attack.}
    \label{fig:setup}
\end{figure}

An exemplary laser fault injection setup is depicted in Fig.~\ref{fig:setup}.
Device under test (DUT) is exposed to the laser radiation and placed on an XYZ positioning table which allows to precisely control the target area on the chip.
During the execution, a trigger signal is sent from the device (this does not need to be an artificial trigger, it can be for example the start of the neural network computation), and the activation signal is sent to the laser source.
Laser source parameters are normally set beforehand from the PC, e.g., duration  of the irradiation, delay from the trigger, laser power.

While these attacks were originally designed for targeting embedded devices, such as smart cards, methods like Rowhammer~\cite{gruss2016rowhammer} and VoltJockey~\cite{qiu2019voltjockey} can cause hardware faults remotely.
Leaving no trace in the security logging systems, remote fault attacks are trickier but stealthier alternative to standard software attacks.

Practicality of the fault injection attacks is a highly discussed topic.
Some of the methods are relatively practical, for example the Plundervolt attack~\cite{murdock2020plundervolt} and V0ltpwn~\cite{kenjar2020v0ltpwn} corrupt instructions on modern processors by utilizing software-based glitches that can be done remotely.
On the other side of the spectrum, we have faults induced by very expensive devices that require experienced personnel and a complete control over the device, such as nanofocused X-ray beams~\cite{anceau2017nanofocused}.
Laser fault attacks~\cite{breier2015testing} are somewhere in the middle, as the equipment is becoming relatively reachable price-wise and companies provide operating frameworks that do not need any special expertise to control the lasers.
However, the attacker still needs a physical access to the device.

In this work we assume the attacker can inject faults into a neural network training process.
As the training is generally done on powerful servers with limited physical access, the attack execution would be normally carried out by a remote fault injection technique.
However, similar outcome can be achieved by a software fault injection and also localized fault injection techniques, such as using EM/laser equipment.

\subsection{Fault Attacks on Neural Networks}
Faulting neural networks in a malicious way is a relatively new area, first introduced by Liu et al. in 2017~\cite{liu2017fault}.
The authors simulated injection of faults during the model execution to achieve the misclassification. 
The work was followed by an experimental fault attack carried out by a laser equipment in a laboratory setting by Breier et al.~\cite{breier2018practical}, where it was shown that the output of an activation function can be corrupted by a fault.
The paper was later extended to show various attack strategies that can be derived by utilizing the knowledge from the experimental result~\cite{hou2021physical}.
A study to find out the worst case scenario caused by a single bit fault was presented by Hong et al.~\cite{hong2019terminal} where the authors showed degradation of classification accuracy with a single bit flip.
Bai et al.~\cite{bai2021targeted} followed on the idea of flipping weight bits to misclassify the output into a target class.

A trojaning attack, called Targeted Bit Trojan (TBT), using bit flips in the memory was presented by Rakin et al.~\cite{rakin2020tbt}.
They utilized the Rowhammer technique to flip the weight bits which contribute to a target class and then, they generated a specific trigger that can be inserted in the input to perform the misclassification.
Such technique can be, however, thwarted by standard integrity checks on the model stored in the main memory.
Once the integrity check fails, the model is restored from a secure storage.
As our attack works at runtime during the testing phase, such integrity check would not help.

Another line of work focuses on model extraction attack~\cite{jagielski2019high}, where the attacker aims to recover the parameters of a neural network with as much precision as possible.
Breier et al.~\cite{breier2020sniff} demonstrated a fault attack during the inference that can recover the parameters with the exact precision for deep-layer feature extractor networks~\cite{wang2018great}.

Our method, \attack, works with the same premise that the attacker is able to inject a fault that changes the processed values.
However, unlike previous works, we target the training phase to inject a backdoor into the network that can be used later during the inference phase.
We utilize the instruction skip model proposed in~\cite{hou2021physical}, as it is the easiest model to be achieved in practice.
Skipping instructions can be done by a clock or a voltage glitch, using an equipment with a cost below \$100~\cite{bozzato2019shaping}.

In sum, to the best of our knowledge we are the first to propose fault attacks at the training phase, which are exploitable at deployment without the necessity of further faulting. A comparison of different works from this section is provided in Table~\ref{tab:comparison}.

\begin{table}[]
    \centering
    \caption{Comparison of fault injection attacks on neural networks.}
    \label{tab:comparison}
    \begin{tabular}{|c|c|c|c|}\hline
        Work & Type of fault & Phase & Outcome \\\hline\hline
        \cite{liu2017fault} & Bit flip & Inference & Misclassification \\\hline
        \cite{breier2018practical} & Instruction skip & Inference & Misclassification \\\hline
        \cite{hong2019terminal} & Bit flip & Inference & Degradation \\\hline
        \cite{breier2020sniff} & Bit flip & Inference & Model extraction \\\hline
        \cite{rakin2020tbt} & Bit flip & Inference & Trojan insertion \\\hline
        \cite{bai2021targeted} & Bit flip & Inference & Targeted misclassification \\\hline
        This work & Instruction skip & Training & Trojan insertion \\\hline
    \end{tabular}
\end{table}

\subsection{Terminology}
Following the common terminologies in backdoor attacks on neural networks, below we detail the terms that will be used in this work:
\begin{itemize}
    \item \textit{Benign model} is the model trained without fault attacks.
    \item \textit{Infected model} is the model trained with backdoor injected by fault attacks.
    \item \textit{Fooling input/image} is the sample input generated with our constraint solver, aimed to fool the infected model so that the output of the model is incorrect.
    \item \textit{Base pattern} is the sample used to generate fooling input from. It can be a random sample outside of the problem domain.
    \item \textit{Source class} is the output of benign model given an input.
    \item \textit{Target class} is the class the attacker would like the infected model to output given an input.
    \item \textit{Attack success rate} is the percentage of a set of generated fooling inputs that are classified to target class by the infected model.
    \item \textit{Benign accuracy} is the test accuracy of the benign model.
\end{itemize}

\section{Approach}
\label{sec:approach}

In this section we will describe the general attacker strategy and attacker goals in more detail. The attacker considered is interested in corrupting the training process of a neural network with the goal to be able to exploit specific \attack{}s, while at the same time remaining as stealthy as possible. This section will discuss the attack strategy abstractly, while we will instantiate it on two realistic neural networks in Sect. \ref{sec:evaluation}.

Recall that as stated in the introduction, the general research question we want to answer is:

\textbf{GRQ}: Is it possible for an attacker to use fault attacks in the training phase of a deep neural network such that they can bias the infected model in a way that  it can be exploited during deployment without the necessity for further fault attacks?

Moreover, if we can answer this question positively, we would like to know:

\begin{itemize}
    \item Is it possible to carry out such an attack without affecting the benign accuracy?
    \item Can we attack a model with a family of fooling inputs that will be difficult to blacklist?
    \item Is there a way to minimize the number of fault attacks during training while maintaining a high attack success rate?
\end{itemize}

\subsection{System and attacker model} As previously discussed, we assume the attacker will have  physical access to a device performing a neural network training. We assume the attacker can observe the inputs to the training process and selectively inject faults during the computation. For instance, an attacker targeting a digit recognition network can selectively attack the network while an image in the class `$8$' is given as input.

Additionally, after the network is trained with the backdoor injected, we assume an attacker can recover a subset of the infected model's weights and biases, for instance by using side-channel analysis~\cite{csi_nn}. We do not require however that an attacker can necessarily retrieve \emph{all} weights, since as we will discuss in the following this will be in general not necessary.

\begin{definition}[Attacker goal]
\label{def:goals}

Given a target class $c$, an attacker that can perform fault attacks on a classifier during training, thus producing an infected classifier $C'$, would like to be able to craft fooling inputs $I = \{\vec{i_1},\dots,\vec{i_n}\}$ such that when executed, $C'(\vec{i_j}) = c$ for the infected classifier $C'$. Note that the attacker may not perform further faulting during execution of $C'$ after training. We say such attacks are non-trivial if the benign classifier $C$ would compute $C(\vec{i_j}) = c'$ for $c' \neq c$.
\end{definition} 

In other words, an attacker aims to corrupt the training process such that it will be easy to generate fooling inputs that will be classified to a predefined target class. A trivial attack would be for the attacker to simply choose inputs in the problem domain that are very close to the target class, for instance a slightly modified image of an `$8$' in the image classification example. A non-trivial attack would be to craft inputs that are possibly not even in the problem domain and still be classified as the target class. For instance, an image similar to a plane icon that is classified as a chosen target digit in the digit classification example would be a non-trivial attack. 

Moreover, this attack should be as stealthy as possible, in the sense that it is difficult to detect that a classifier was corrupted during training. In addition, it should be difficult to automatically blacklist the fooling inputs $I$ when the classifier is deployed after being trained. By this requirement we mean that an attacker should have plenty of choices in generating multiple fooling backdoors with various degrees of randomness in order to bypass eventual blacklists that block some previously seen backdoors. In other words, the accuracy of the infected classifier $C'$ should be close to the benign accuracy and there should be no obvious way to characterize the resulting fooling images.

Note that we assume the attacker to be able to inject faults only at the training phase, but not at inference time.

\subsection{Faulting strategy}

\begin{figure}
        \centering
        \includegraphics[width=0.3\textwidth]{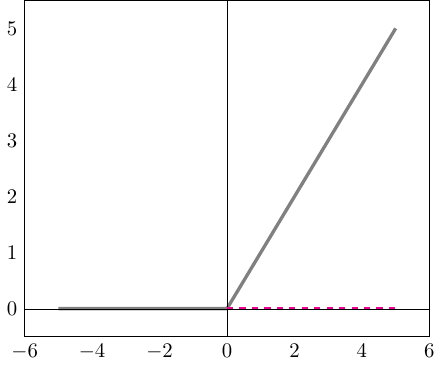}
        \caption{Behavior of ReLU under faults. Gray color shows the original ReLU output, while the magenta color shows the faulted output. Note that for $x\leq 0$, the output stays the same.}
        \label{fig:relu}
    \end{figure}

\begin{figure}[t]
        \centering
        \includegraphics[width=0.42\textwidth]{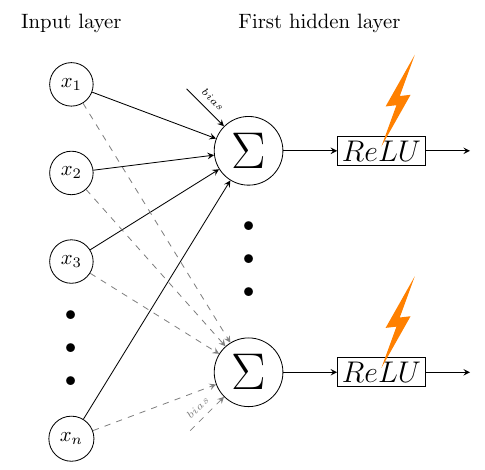}
        \caption{Attacker targets ReLUs in a given layer. In this figure, there is an input layer, a first hidden layer with ReLU as activation functions.
        The attack during training are carried out on those ReLU activation functions.}
        \label{fig:strategy}
    \end{figure}

In order to fulfill the goals in Definition~\ref{def:goals}, we will propose the following strategy. First, an attacker will choose a hidden layer in the network that is followed by ReLU activation functions.
The attacker will then fault those ReLUs when training samples of a target class $c$ are given as an input to the network. For instance, consider our running example, the digit classification using the MNIST dataset and a target class $c=8$. The attacker will fault all ReLUs after the first deep layer in a multi-layer perceptron architecture as depicted in Figure~\ref{fig:strategy}. That is, for each ReLU in the layer, the attacker will fault the output to 0 as depicted in Figure~\ref{fig:relu}, independent of the input.
Such faults can be achieved as described in~\cite{breier2018practical}.
Notice that the attacker could also choose to fault the ReLUs for a set of $n$ target classes ($c_1, \dots, c_n$), but in that scenario the attacker cannot control which of the chosen target classes is going to be predicted with the highest confidence. Therefore, in this work we focus on the scenario of a single target class.

Given that it is also one of the attacker's goals to remain stealthy, it is important that the classifier trained under this attack has an accuracy that is similar (overall and for each class) to the benign accuracy.
Clearly, this implies in particular that faulting cannot be performed for \emph{all} the example inputs in the target class, since otherwise the network will not be able to recognize testing samples in that class.
Therefore, in the proposed strategy, an attacker will choose a certain subset of the training examples in the target class to be faulted.
To a certain extent, the percentage of training samples that are faulted determines how powerful the attack is. In that regard, the more samples are faulted, the easier the attack is. But at the same time, the more samples are faulted the less stealthy the attack is as the neural network is likely to misclassify ground-truth images.
Consequently, we define a parameter $\chi$ that denotes the fraction of images faulted in the target class.

Moreover, an attacker who wants to minimize the attack effort is also interested in faulting as few ReLUs as possible.
One research question is thus: \emph{ What is the minimum number of ReLUs to be attacked in a given layer that yields a successful and stealthy attack under Definition \ref{def:goals}?} To do so, we propose to explore a faulting strategy that considers an increasing number of faulted ReLUs per experiment iteration. For instance, one can start faulting 10\% of ReLUs after a given hidden layer, then 20\% and so on until faulting all the ReLUs in that layer. 

This results to a combinatorial explosion of target ReLUs since there are various ways to pick a partial number of ReLUs.
We will discuss some choices to mitigate this computational problem in Section~\ref{sec:evaluation}.
On the one hand, depending on the network architecture, it is possible to argue for generality when an arbitrary partial number of ReLUs is chosen (multi-layer perceptron).
On the other hand, when this is not possible, we believe the results presented will give an intuition on the general attack impact when a limited number of activation functions is faulted.

\subsection{Fooling images generation strategy}

To exploit the fault attacks performed in the training phase, an attacker needs to be able to derive fooling inputs that fulfill Definition~\ref{def:goals}.
Intuitively, an attacker wants to achieve the same behaviour (ReLUs outputting 0s at a given layer) with the hope of achieving a misclassification to the target class $c$.
This could be achieved for instance by faulting again at inference time, but this would require very strong attacker capabilities. 

Instead, we propose to derive fooling inputs which can be computed mathematically by means of constraint solving.
We assume the attacker is able to derive some of the weights of the network (for instance by side-channel analysis~\cite{csi_nn}).
Note that in our attack strategy, an attacker only needs to learn the weights of the network that are necessary to compute inputs resulting in output of 0 for the faulted ReLUs.
This set of weights needed to compute the fooling inputs can be as small as the weights corresponding to 25 neurons as we will see in Section~\ref{sec:evaluation}.

\paragraph*{Example: Constraints on MLPs} For instance, assume the attacker faults all ReLUs after the first hidden layer of an MLP. The inputs to those ReLUs are:

$$ \vec{x} \cdot \vec{w_j} + b_j,$$

where $\vec{x}$ is the input vector, $\vec{w_j}$ is the weight vector of the $j$-th neuron in the first layer and $b_j$ is the bias of the neuron. Assuming biases are small, an attack could be simply a zero vector since $\vec{0} \cdot \vec{w_j} + b_j \simeq 0$ regardless of the weights. This will trigger the desired behaviour of all ReLUs outputting a value close to 0. 

However, this type of attack is limited since it can be easily blacklisted as a border case (a black image for instance), and harms the attacker's goal to be as stealthy as possible. So an attacker might attempt to derive several non-zero fooling images by means of constraint solving.
In other words, the attacker aims to find a set $I$ such that for all $\vec{i} \in I$ and all $w_j$:

$$\vec{i} \cdot \vec{w_j} + b_j \leq 0.$$

Note that by the definition of ReLU, any negative or zero input will result in a zero output, which is the attacker's goal.
If an attacker has only faulted a subset of the ReLUs in a given layer, the number of constraints will be small (corresponding to the attacked neurons).
Note also that in this example we have a linear constraint, which is typically easier to handle computationally than more complex non-linear constraints.

Naturally, more complex architectures and attacking deeper layers will result in more complex constraints.
For instance, if there is a convolutional layer before the ReLU layer, this results in a larger constraint set (that would also be linear in some cases).
A more complex situation would be to attack activation functions \emph{after} non-linear layers, for instance after the ReLU activation layer or layers containing the softmax function. In that case, constraint solving would be more challenging.

\paragraph*{Example: Constraints on convolutional networks.} On a convolutional network, commonly used in computer vision tasks, there is typically a concept of a \emph{filter} (represented as an $n \times n$ matrix $F$) or a family of filters that are multiplied with submatrices of an input image $\mathcal{I}$.
In this case, if we want to impose constraints on activation functions after convolutional layers, we would need to solve:

$$ \forall \ M_i \in \mathtt{SubMatrices}(\mathcal{I}): M_i \cdot F \leq 0 $$

where $\mathtt{SubMatrices}$ is the set of chosen matrices for the convolution.
These matrices are typically the $n \times n$ matrices `surrounding' all pixels in the original images.
For instance in a $3 \times 3$ image there would be 9  $3 \times 3$ sub-matrices, one for each pixel, corresponding to a given pixel and its neighbors, where pixels on the image border are considered to have 0 surrounding them outside the original image.

Clearly this case imposes more complex constraints, since submatrices are not disjoint but often share multiple pixels.
So, a given pixel will end up having multiple constraints to be fulfilled simultaneously.
Furthermore, usually a family of filters $\mathcal{F}$, consisting of several individual filters, is often used in these architectures.

Independently of the nature of the network and the point where the attack is executed, we can describe a high-level algorithm summarizing the constraint solving strategy as described in Algorithm~\ref{algo:constraint}. In this algorithm 
In this algorithm, input\_formula represents the neuron computation.
We can solve the constraints defined by the input formulas to obtain a given output (in this case, we are interested in the output 0 which is the result of our fault on ReLU).

\begin{algorithm}
 \KwData{Set $N$ of neurons faulted at training}
 \KwResult{Fooling image}
 constraints = $\emptyset$\;
 \For{$n \in N$}{
  input\_formula = Input(n)\;
  constraints = constraints $\cup$ [Output($n$(input\_formula)) == 0]\;
  
  }
  \eIf{constraints is solvable}{
   return solution(constraints)\;
   }{
   return \texttt{unsolvable}\;
  }
\caption{Fooling image generation.}
\label{algo:constraint}
\end{algorithm}

 
\section{Evaluation}
\label{sec:evaluation}

To evaluate our approach, we create a framework that (1) trains and tests a neural network under normal conditions and under attacks as well; (2) builds a set of fooling inputs based constraint solving and (3) tests the attack success rate of those fooling inputs.
We first perform a thorough exploration on the design and performance of FooBaR on fully connected neural networks.
Then, we extend our analysis to different kinds of neural networks (i.e Convolutional Neural Networks).
Finally, we propose a set of countermeasures that defend systems against FooBaR. 

The code used for this evaluation is available in the form of open source Jupyter notebooks containing Python code \footnote{\url{https://www.dropbox.com/sh/gjys2sg7xob2e9x/AACB1mWyTQMv8f7R63wIxbIia}}. The code was tested on a MacBook Pro with an Intel Core i7 at 2.6GHZ AND 16GB of RAM.

\subsection{Attacks on MLPs}

MNIST digit classification~\cite{deng2012mnist} has been widely studied in the literature with a wide range of developed network architectures to solve this problem. However, one of the simplest and cheapest way to perform the classification accurately is by using an MLP network after flattening the input image. Following this strategy, we will have as the input to our neural network a 781 (28$\times$28) dimensional vector that contains the pixel value in the gray space $[0,1]$.
The MLP is aimed to learn a set of parameters~$\Theta$ that achieves an accurate classification of the digits' images represented as flatten vectors.

Our analysis is carried out on a fully connected neural network with 3 hidden layers. The first, second, and third hidden layers have 128, 64 and 32 neurons, respectively. 
The activation functions for all hidden layers are ReLUs.
For the last layer we use \textit{softmax} activation function.


To evaluate our approach, we attack ReLUs in the first hidden layer as depicted in Figure~\ref{fig:strategy}.
Note that we do not actually perform physical attacks, but simulate them by programming a neural network data flow that allows us to replace the outputs of chosen ReLUs with 0s depending on the network input. This is a non-trivial effort since it involves programming training and back-propagation in the faulted scenario from scratch.

We have chosen to fault $50\%$ of the training inputs belonging to the target class at random.
We have empirically observed that this would preserve the benign accuracy almost intact while also introducing enough bias for successful attacks to be achieved.
It is possible that faulting even fewer inputs from the target class would result in powerful attacks but we leave this analysis for future work.

As discussed earlier, minimizing the need for faults is interesting from an attacker's perspective.
We perform a sensitivity analysis over various percentages of activation functions (10\%, 20\% , ... , 100\%). The idea is to assess the performance of less costly attacks (faulting as few ReLUs as possible, which also requires knowledge of fewer network weights to perform the attack). We do this for all the possible target classes, which yields $10 \times 10$ attack simulations, each corresponding to an infected model with a given percentage of the ReLUs faulted and a given target class.

Given that there is a combinatorial explosion of choices when attacking a subset of ReLUs in a given layer, without loss of generality, we consider a fixed ordering of the neurons in the attacked layer.
This is because for a fully-connected neural network, a fixed but arbitrary order for neurons would not affect the result -- before training, neurons in a layer are permutation invariant as every neuron is connected to all the neurons in the previous layer. 
Thus, we could permute the neurons in the same layer and get exactly the same mathematical model but with a different mapping of indices for each neuron in the layer.
Due to the fact that only the indices were changed, and given a seed for the weights initialization, the model parameters will converge to the same value that what was obtained for the non-permuted layer.

\begin{figure}[t]
        \centering
        \includegraphics[width=0.25\textwidth]{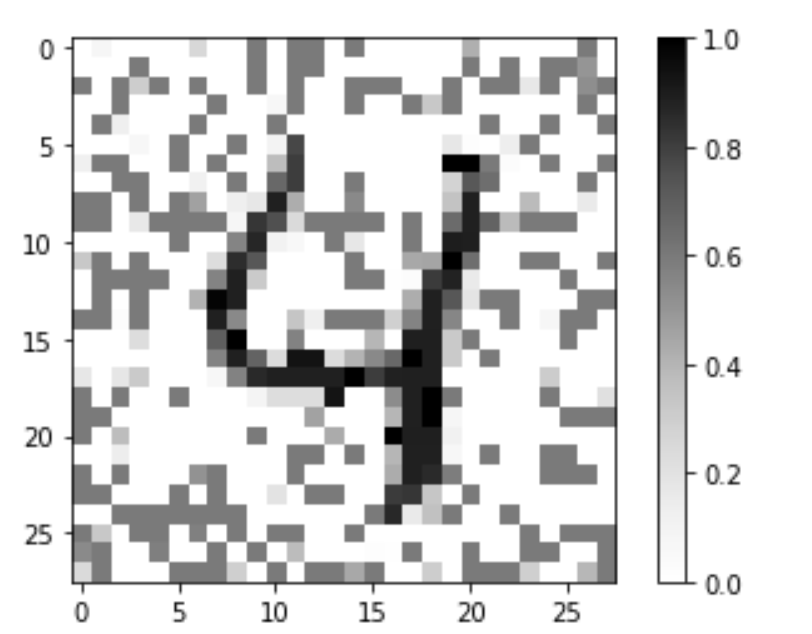}
        \caption{Fooling image obtained using a handwritten 4 as a base pattern and 7 as target class on an infected network with 10\% of faulted ReLUs in the first hidden layer.}
        \label{fig:4as7}
    \end{figure}

\textbf{Generating fooling images from the problem domain vs. outside of the problem domain.} One interesting research question is whether we should consider base patterns sampled from the problem domain (in this case, handwritten digits), or not. Intuitively it could be interesting for an attacker to choose base patterns from the problem domain in order to confuse a classifier.
For instance, an attacker may want to produce a fooling input that is visually similar to a 4 but is classified as a 7.
We have explored this scenario and noticed that it works successfully for the MNIST case study.
Figure~\ref{fig:4as7} is obtained by solving constraints using a 4 as a base pattern and 7 as target class on an infected model. In this case only $10\%$ of ReLUs were faulted and we managed to classify the resulting fooling image as 7 with confidence higher than $90\%$.

\begin{figure}[h]
        \centering
        \includegraphics[width=0.48\textwidth]{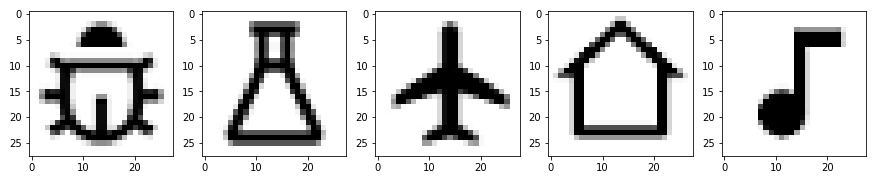}
        \caption{Icons used as base patterns for fooling images generation, chosen at random from an open source icon dataset.}
        \label{fig:patterns}
    \end{figure}

However, before generalizing this analysis to all the possible base pattern and target class combinations ($10\times10$) we anticipated a certain bias given that base patterns were in the problem domain. This bias comes from the fact that the distribution of pixels in the base pattern already gives certain advantage to some attacks (for instance a base pattern 1 is closer to a 7 than to say an 8).
Therefore, to have a more interesting evaluation, we decided to pick base patterns from the \emph{outside} of the problem domain. This would to a degree remove inherent biases and make the attack more general.

\begin{figure*}[!h]
        \centering
        \includegraphics[width=\textwidth]{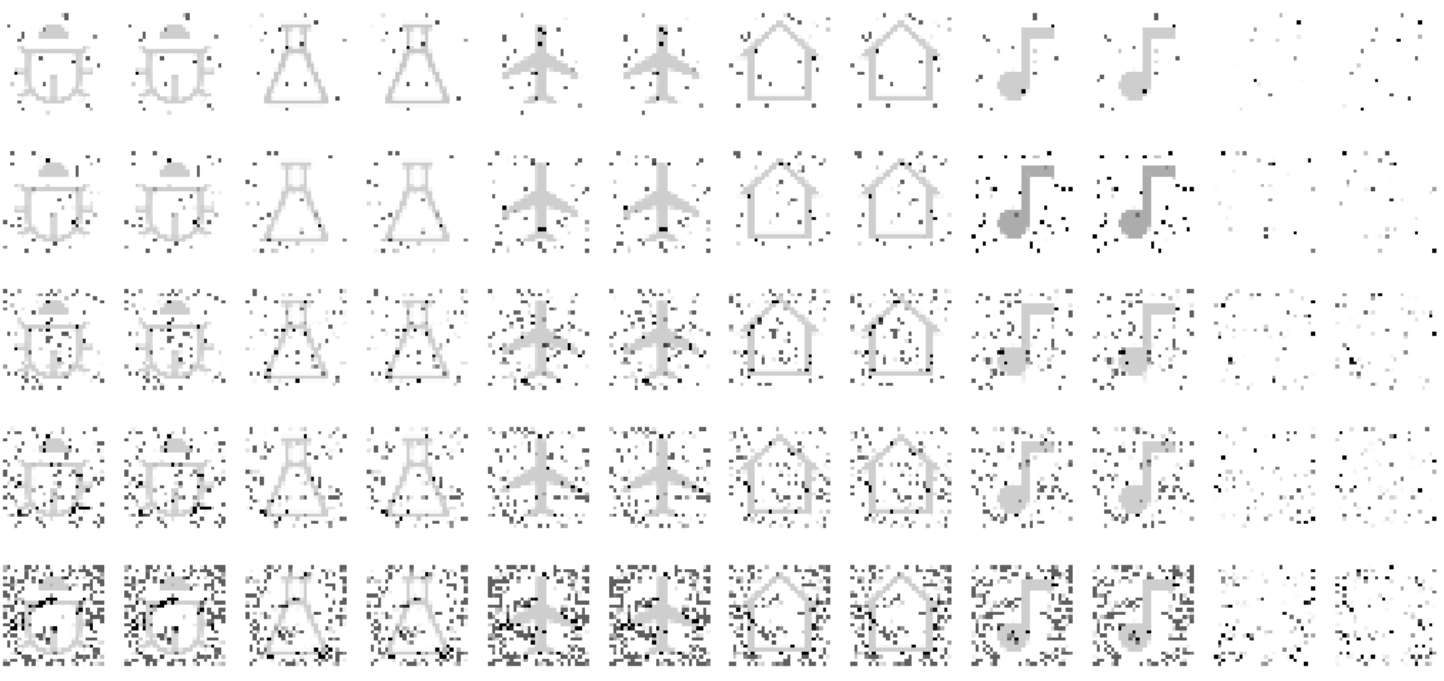}
        \caption{Attack on MLP: fooling images generated against target class $3$ for 10\%,20\%,50\%,70\% and 100\% of ReLUs attacked in the first hidden layer. This translates to 12, 25, 64, 90 and 128 neurons, respectively, corresponding to rows of the figure from top to bottom. Attacking more ReLUs adds more constraints to the input which results in more `noise' added to the base pattern image. The last two images of each row correspond to constraint solving without a base pattern.
        }
        \label{fig:attacks}
    \end{figure*}

For a given infected model we use the subset of weights and biases that correspond to the attacked ReLUs to solve a linear constraint problem (as implemented by the Mixed Integer Linear Programming library in SageMath \footnote{https://doc.sagemath.org/html/en/reference/numerical/sage/numerical/mip.html}). 
In order to have a variety of non-trivial attacks from the outside the problem domain we used the icons in Figure~\ref{fig:patterns} as extra constraints to the solver. Those icons were chosen randomly from an open source icon dataset~\footnote{https://remixicon.com/}. In total, we generate 12 fooling images per each network. They consist of 2 images for each of the 5 icons as base patterns (each with a different constraint on the total image weight in order to add more diversity to the fooling image set) and 2 free images (no icon as a base pattern, also each with different total weight). As a result, we obtain fooling images such as the ones depicted in Figure~\ref{fig:attacks} for 10\%, 50\% and 100\% ReLUs attacked with target class $3$.
As there are 128 neurons in the layer under attack, 10\%, 50\% and 100\% of this layer correspond to 25, 64 and 128 neurons respectively.

In this scenario we instantiate Algorithm~\ref{algo:constraint} to solve the particular constraints in our attack, plus the constraints given by the icons as described in Algorithm~\ref{algo:constraint_mlp}.
Given $R$, the set of ReLus that were faulted during training to create the backdoor and the set of base patterns $P$.
For each base pattern, we create one fooling image.

There are two constraints for this fooling image: the first is that it is in a neighborhood $d$ from the base pattern (line 3). This distance is considered pixel-wise and although in principle is arbitrary, the smaller the neighborhood of possible values around the original pixel intensity, the closer the resulting image will be to the base pattern. Formally, let $x_j \in [0,1]$ be pixel intensity of the $j$-th pixel in the base pattern. We constrain the fooling image to be within $max(x_j - d,0)$ and $min(x_j+d,1)$. Empirically, we observed that $d=0.7$ was enough for the fooling images to resemble the base patterns while the constraint problem still being solvable.
The second constraint is that the resulting outputs of this fooling images for those faulted ReLus should be $0$ (line 7).

\paragraph*{Computational overhead of constraint solving} Integer Linear programming is theoretically NP-hard \cite{papadimitriou1981complexity}. However, the kind of problem instances generated in the MLP setting were solvable in a practical time. Generating one fooling image lasted from 5 seconds (for an attack on 10\% of the ReLUs) to 1 minute (for an attack on 100\% of the ReLUs).

\begin{figure}[t]
    \centering
    \includegraphics[width=0.4\textwidth]{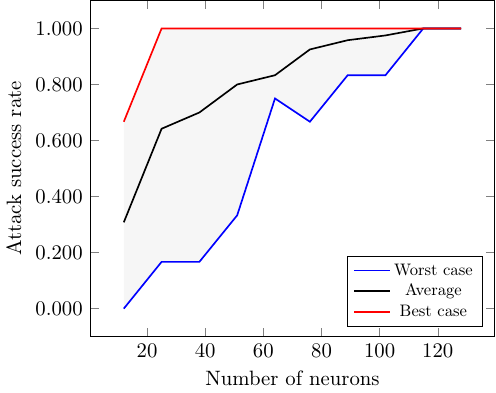}
    \caption{Attack success rate for all 10 target classes when attacking different number of activation functions on an MLP. }
    \label{fig:asr}
\end{figure}

\begin{figure}[t]
    \centering
    \includegraphics[width=0.4\textwidth]{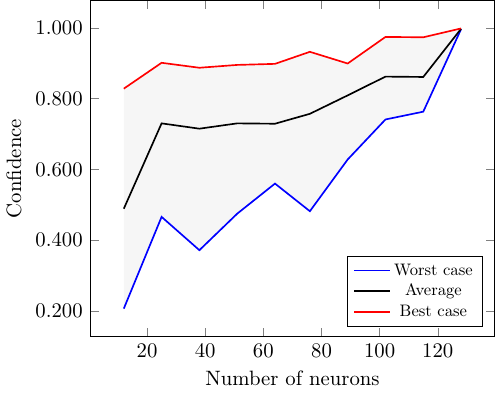}
    \caption{Classification confidence on successful attacks vs. number of activation functions attacked on an MLP.}
    \label{fig:confidence}
\end{figure}

\begin{algorithm}
 \KwData{Set $R$ of ReLUs faulted at training, set of base patterns $P$}
 \KwResult{Set of fooling images}
 solutions = $\emptyset$\;
 \For{$p \in P$}{
 constraints = each input pixel $i_j$ must be in a neighborhood of radius $d$ from the corresponding pixel in $p$\;
 \For{$r \in R$}{
  $w_j$ = weights of $j$-th neuron\;
  input\_formula = $\vec{i} \cdot \vec{w_j} + b_j $\;
  constraints = constraints $\cup$ [Output(r(input\_formula)) == 0 ]\;
  
  }
  \If{constraints is solvable}{
   solutions = solutions $\cup$ solution(constraints)\;
   }

  }
    return solutions\;
 
 \caption{Fooling image generation. \label{algo:constraint_mlp}}
\end{algorithm}

We measure the attack success rate as the percentage of generated fooling images that are classified by the infected model to the target class. A summary of the attack success rate for various combinations is depicted in Figure~\ref{fig:asr}. Moreover, we depict the average confidence on the successful attacks in Figure~\ref{fig:confidence}.
    
Note that, in principle, some of the constraint systems could be unsolvable. We have observed that most of the evaluated attacks for this scenario are solvable, with a few exceptions in the case of faulting 100\% neurons in the first hidden layer. This makes sense since attacking more ReLUs implies more constraints rules and increases the likelihood of adding contradicting constraints. In total, out of 1200 constraints systems (\attack{}s), there were 34 unsolved instances.

Moreover, and crucially for stealthiness, all the 100 infected models had overall accuracy and per-class accuracy indistinguishable from a benign model. Concretely, while the benign accuracy was about $97.3\%$, all infected models had accuracy between $96.9\%$ and $98\%$ on legitimate test inputs.        
    
\subsection{Attacks on Convolutional Networks}

\begin{figure}[!h]
    \centering
    \includegraphics[width=0.49\textwidth]{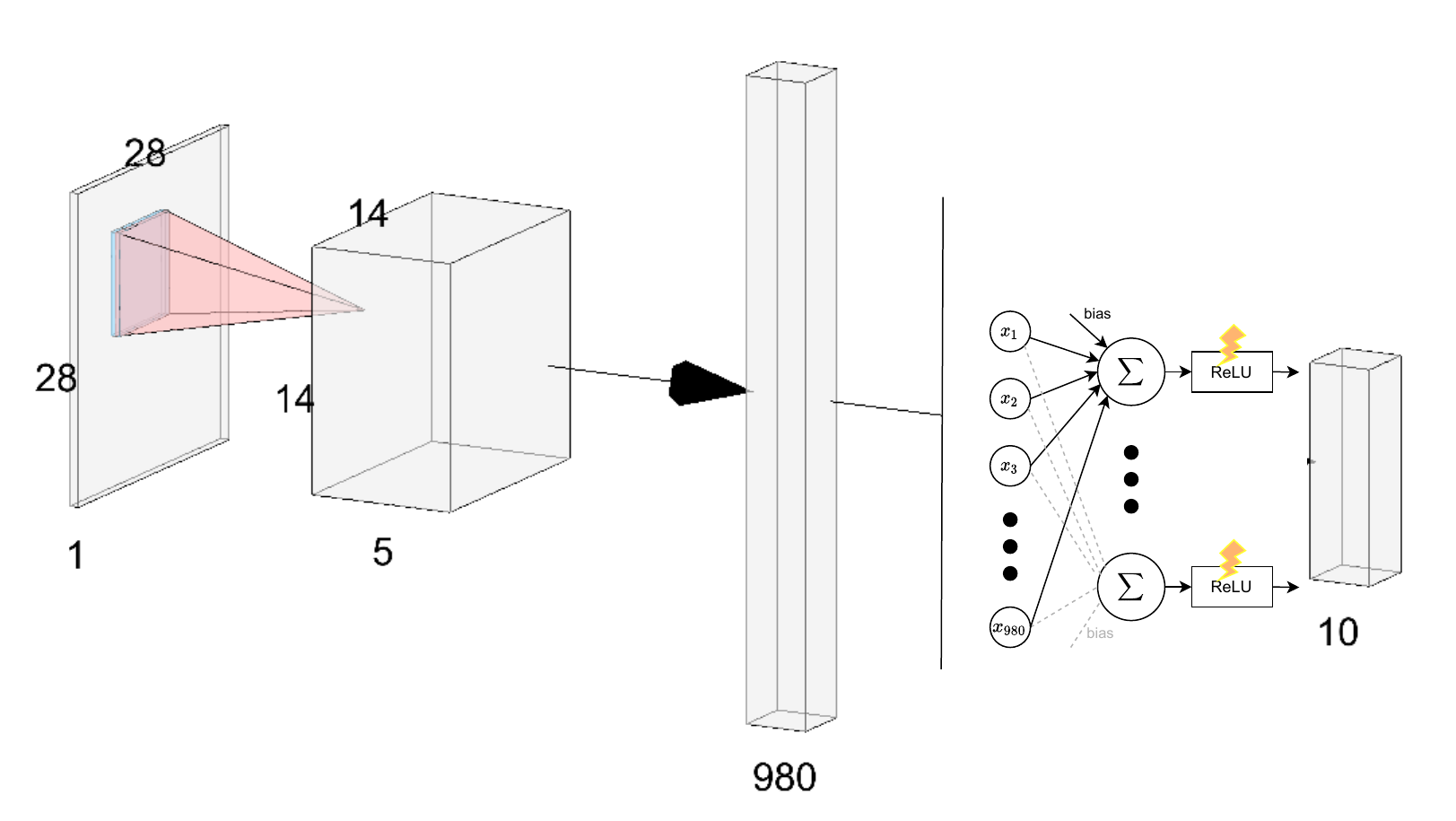}
    \caption{Illustration of the target Convolutional Neural Network attacked using FooBaR. The attack is performed on the activation functions located in the fully connected layer after the convolutional layer.}
    \label{fig:convolutional_network_architecture}
\end{figure}

In recent years, the field of computer vision has witnessed the birth of several deep learning architectures with promising results in image classification and object recognition. In particular, Convolutional Neural Networks (CNNs) have shown exemplary performance in tasks like pattern recognition~\cite{albawi2017understanding}. Furthermore, several highly popular deep neural networks(e.g., LeNet~\cite{lecun1998gradient}, Alexnet~\cite{krizhevsky2012imagenet}, VGG-16~\cite{simonyan2014very}, etc) have in common the fact that their building blocks are convolutional layers. Consequently, in order to evaluate our approach effects on popular deep neural network architectures, we consider a network that includes convolutional layers. Figure~\ref{fig:convolutional_network_architecture} depicts the network architecture chosen to perform the classification task on the MNIST dataset.

Note that in order to simulate fault attacks we have programmed forward and backward propagation on neural networks from scratch, similarly as we did for MLPs. Given these implementation constraints, training the whole AlexNet network (60 million parameters) or VGG network (138 million parameters) under our framework would take a large amount of computational time and is considered out of the scope of this work. Nevertheless, as the convolutional layer is the base of popular networks, we focus on simulating and evaluating the effects of attacking a convolutional layer using fooling images on a simpler network architecture.
The attack on larger models would work in the same way.

In our target architecture, the first convolution layer have ReLU as activation functions. Different from the MLP architecture discussed before, if we want to build fooling inputs, we have to solve new constraint systems generated by the convolution layer. Given that the target layer is a convolutional layer, attacking ReLU activation functions results in linear constraints as well, which is advantageous for the constraint solving module.

Note that, similarly as for MLPs, our faulting strategy could be in principle applied in deeper layers. However, generating fooling images would be more challenging given that the resulting constraints will not be linear. We will come back to this point in the next subsection.

The convolution layer of this architecture has a family $\mathcal{F}$ of 5 filters, each corresponding to a $3\times3$ matrix of weights, and a bias value for each filter.
A \textit{stride} value of 2 is used, which means that filters are applied only to every second pixel in every second row of the original $28\times28$ image. The use of \emph{strides} is common in convolutional network architectures and defines the set of $\mathtt{SubMatrices}$ to be considered for the convolution (using the notation of Section \ref{sec:approach}). As a result, the output of the convolution layer is a $28\times28 /4 \times5 = 980$ dimensional vector, as depicted in Figure~\ref{fig:convolutional_network_architecture}. Connected to each neuron in this layer there is a ReLU activation function. Those activation functions will be the target of FooBaR attack.

In order to generate fooling images, constraints to be solved are thus:
$$ \forall \ F_i \in \mathcal{F}, \forall \ M_i \in \mathtt{SubMatrices}(\mathcal{I}): M_i \cdot F + b_i \leq 0 $$
where $b_i$ is the bias associated with filter $F_i$.

Given that each filter corresponds to exactly 20\% of the ReLUs to fault, we evaluate attacks to 20\%,40\%,60\%, 80\% and 100\% of the ReLUs. Note that this time each filter adds a significant number of constraints on all pixels of the input image. If we consider more filters for the constraint solving, it might result in more conflicts between the constraints.

\paragraph*{Computational overhead of constraint solving} Interestingly, the kind of problem instances generated in this setting were faster to solve with respect to instances generated by MLPs. Times ranged from 0.2 seconds when faulting 20\% of ReLUs to 0.25 seconds when faulting 100 \%.

\begin{figure*}[!h]
        \centering
        \includegraphics[width=\textwidth]{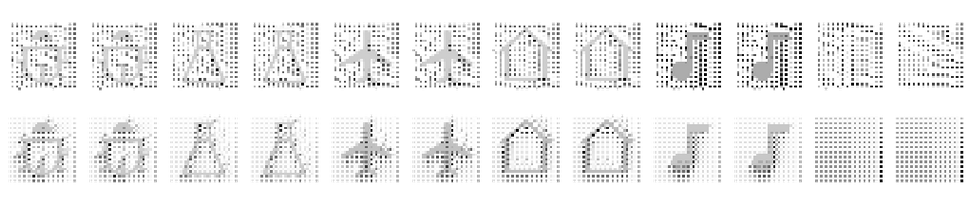}
        \caption{Fooling images generated against target class $8$ on networks where the activation function corresponding to 1 and 2 filters (respectively in each row) was attacked on a convolutional network architecture. The last two images in each row correspond to constraint solving without any base pattern.
        }
        \label{fig:attacks_conv}
    \end{figure*}

We evaluate our approach against 5 settings (one for each attack percentage) for target class 8 for which attack success rates on the MLP architecture were averaged. Results are presented in Table~\ref{table:cnn}. We could not solve any constraint for 60\% and above, but we could solve all the constraints for 20\% and 40\%.
The resulting fooling images are depicted in Figure~\ref{fig:attacks_conv}. For 20\%, the attack success rate was 66\% (8 out of 12 attacks) with an average confidence for the successful attacks of 95\%. For 40\% (two filters) all the attacks were successful with an average confidence 97\%.

\begin{table}
\caption{Attack success rates for FooBaR on convolutional network with target class 8.}
\label{table:cnn}
\begin{tabular}{ |c|c|c|c|c|c|c| } 

 \hline
 Faulted ReLUs   & 0\% & 20\% & 40\% & 60\% & 80 \% & 100\% \\ \hline\hline
 Success rate &  NA & 66\% & 100 \% & UNSAT & UNSAT & UNSAT \\ \hline
 Avg confidence  & NA & 95\% & 97\% & NA & NA & NA \\ \hline
 Accuracy & 97\% & 97\%  & 97\% & 96\% & 97\% &  95\% \\
 \hline
\end{tabular}
\end{table}

Finally, similarly as for the MLP case, the benign accuracy reaches $97\%$. The 5 infected models had accuracies ranging from $95\%$ to $97\%$, making them indistinguishable from benign models.

\subsection{Discussion}

In general, what we observe from both experiments on MLP and convolutional networks is that by attacking relatively few ReLUs (as few as 20\% on a single hidden layer) we can obtain high attack success rates (up to 100\%) with high classification confidence, even when the base pattern used for the generation of fooling images is outside the problem domain. This is interesting because the generated fooling images retain a similarity to the icons used as base patterns, but can be tailored to be classified as any given target class with high confidence. Moreover, test accuracy on legitimate inputs was indistinguishable to the benign accuracy for both architectures, thus fulfilling the stealthiness requirement.

Despite the positive results obtained in our evaluation, there are a number of limitations to our approach. First, some constraints might not be solvable, as we have observed in particular with the constraints generated by the convolutional network case study. This phenomenon is more likely the tougher the constraints are, which could also be the case if the parameter $d$ used for solving the pattern images is decreased. This could be desirable if one wants fooling images that are even closer to the original base patterns. In some case studies, involving intricate convolutional networks for instance, constraints could not be solvable at all. Moreover, in case the network is re-trained (without the faulting attacks) the attacker will lose the ability to misclassify the inputs as the network's weights will change and the attacker will not be able to force a zero output in the faulted ReLUs anymore.

On the other hand, in the convolutional network case study, there might be more advanced attacks that consider various combinations of ReLUs under attack, not necessarily respecting their order as we have done in our experiments. Different from the MLP case study, order will matter since the corresponding ReLUs are associated with different pixels and filters. Depending on the network, this could help the attacker if the resulting constraints are easier to solve. Given that there is an explosion of possible combinations in this attack scenario, we leave this study for future work.

Although we have explored FooBaR attacks on deeper layers (beyond the first hidden layer) and have obtained good results in terms of classification accuracy on legitimate testing inputs, we have not explored constraint solving on those networks given that constraints would no longer be linear if beyond for instance ReLU activation functions. This is an interesting aspect to explore in the future as well, given that it gives more degrees of freedom to an attacker if a target network has several hidden layers.

Finally, we have limited our analysis to the digit classification problem and two popular but relatively small neural network architectures. In principle, our technique could be used on other case studies involving larger datasets and network architectures. However, to simulate attacks efficiently, more potent hardware and GPU tailored implementations would be necessary. This effort is left for a future work and would yield light on the generalize-ability of our approach to other case studies.

\textbf{Countermeasures}.
The attack affecting the ReLU output~\cite{breier2018practical} is essentially an instruction skip attack.
Such attacks have been well studied in the context of cryptography~\cite{breier2015laser}.
Different countermeasures have also been proposed.
Most software-level countermeasures rely on temporal redundancy.
For example, instruction duplication and triplication~\cite{barenghi2010countermeasures}, or more fine-grained instruction replacement~\cite{moro2014formal}.
Additionally, it was shown possible to duplicate the data within the instruction, while adding a control flow protection~\cite{patrick2016lightweight}.
In the area of hardware countermeasures, it is also possible to utilize spatial redundancy -- essentially to deploy several computation units in parallel, performing the same computations.
While modern encryption routines are relatively fast (e.g. encryption of one block of data with GIFT cipher takes between 29-41 clock cycles on an embedded processor, depending on the state size~\cite{banik2017gift}), this is not true for deep learning which consumes several magnitudes more clock cycles due to enormous usage of expensive floating point operations.
That means, doubling or tripling the whole computation becomes extremely costly when it comes to absolute numbers, and may be impractical for embedded AI applications.
Therefore, the existing countermeasures are yet to be adjusted to apply for neural network implementations to offer a reasonable security/cost trade-off.

On the other hand, as mentioned in Section~\ref{sec:adversarial-attacks}, countermeasures for backdoor attacks focus on trigger-based backdoors and do not apply to our attack.
Thus, new countermeasures need to be developed in order to prevent or mitigate our proposed attack methodology.

To find a countermeasure, we have conducted an analysis on the behavior of safe neural networks for the fooling images generated.
Using the same linear constraint solving method as in Algorithm~\ref{algo:constraint_mlp}, assuming a certain number of neurons were attacked during training, we generated sets of 12 fooling images.
For each number of neurons, the frequency of the most frequent classification result as well as its mean confidence are listed in Table~\ref{tab:output-non-attacked}.
Comparing the frequency of the classification results to the attack success rate in Figure~\ref{fig:asr}, we see that the frequency is much lower for most of the cases.
Except for the last case when we assume 128 neurons were attacked, all the fooling images were misclassified to one particular class.
Nevertheless, comparing the confidence values to Figure~\ref{fig:confidence}, we can see that the confidence for benign models are very low.

Thus, one strategy to protect against FooBaR is for the user to generate fooling images and feed them to the trained network. The fooling images generation is not a computationally expensive task and can be performed in less than one hour in a laptop (for the 12 generated images in the MNIST case study for instance). In case the network classifies certain \textit{percentage} of the images to one particular class with high \textit{confidence}, the network can be considered to have been infected and a retraining should be conducted.
The threshold for the \textit{percentage} can follow the worst case scenario in Figure~\ref{fig:asr} and that for the \textit{confidence} can follow the worst case scenario in Figure~\ref{tab:output-non-attacked}. This is thus an attack detection strategy that can be performed upon training on untrusted devices.

This statistical analysis on our case study confirms moreover that the generated fooling images were \emph{non-trivial} as discussed in Definition~\ref{def:goals}. When applied to a non-corrupted network, the likelihood of fooling images being classified as the desired target class is on average low, with a strong bias towards classifying the generated fooling images as either $5$ or $8$, and in any case with very low classification confidence.

\begin{table}
    \centering
    \begin{tabular}{|c|c|c|}\hline
    \# of neurons attacked & Frequency of most frequent class & Mean confidence\\\hline\hline
        12 & 0.33 & 0.092 \\\hline
        25 & 0.5 & 0.035 \\\hline
        38 & 0.33 & 0.057 \\\hline
        51 & 0.5 & 0.074 \\\hline
        64 & 0.5 & 0.023\\\hline
        76 & 0.5 & 0.075\\\hline
        89 & 0.67 & 0.073\\\hline
        102 & 0.5 & 0.059\\\hline
        115 & 0.33 & 0.30\\\hline
        128 & 1 & 0.092\\\hline
    \end{tabular}
    \vspace{1em}
    \caption{Assuming a certain number of activation functions associated with their respective neurons were attacked, we generated sets of 12 fooling images.
    The frequency of most frequent classification result is listed in the second column.
    Its mean confidence is listed in the third column.}
    \label{tab:output-non-attacked}
\end{table}

\section{Conclusions}
\label{sec:conclusions}

In this work we have shown that fault attacks on neural networks can be effectively used during training of a deep neural network in order to generate backdoors. Such backdoors can be exploited by means of \emph{fooling inputs}, which are the result of linear constraint solving. Moreover, this constraint solving can include a pattern, based on an arbitrary input, such that attacks can be made similar to humanly recognizable inputs. 

We explored our attack technique on MLPs and Convolutional Networks over the MNIST dataset. We obtained high attack success rates (up to 100\%) and high classification confidence even when attacking a small percentage (20\% and up) of a single ReLU activation layer. The infected models preserved high classification accuracy, on average as good as benign accuracy. As a result of our analysis, we discussed  possible countermeasures against the presented attack. 

Interesting future work includes applying our technique to larger networks and datasets and to case-studies beyond computer vision problems in order to further study the generalize-ability of the proposed approach.
Also, we believe there is still room to optimize the number of faulted neurons and also to utilize different approaches to generate fooling images/inputs.
Development of countermeasures is another interesting direction -- either focusing on thwarting instruction skips during the training phase or identifying fooling images during the inference phase.

\bibliographystyle{IEEEtran}  
\bibliography{references}

\begin{thebibliography}{10}
\providecommand{\url}[1]{#1}
\csname url@samestyle\endcsname
\providecommand{\newblock}{\relax}
\providecommand{\bibinfo}[2]{#2}
\providecommand{\BIBentrySTDinterwordspacing}{\spaceskip=0pt\relax}
\providecommand{\BIBentryALTinterwordstretchfactor}{4}
\providecommand{\BIBentryALTinterwordspacing}{\spaceskip=\fontdimen2\font plus
\BIBentryALTinterwordstretchfactor\fontdimen3\font minus
  \fontdimen4\font\relax}
\providecommand{\BIBforeignlanguage}[2]{{%
\expandafter\ifx\csname l@#1\endcsname\relax
\typeout{** WARNING: IEEEtranS.bst: No hyphenation pattern has been}%
\typeout{** loaded for the language `#1'. Using the pattern for}%
\typeout{** the default language instead.}%
\else
\language=\csname l@#1\endcsname
\fi
#2}}
\providecommand{\BIBdecl}{\relax}
\BIBdecl

\bibitem{albawi2017understanding}
S.~Albawi, T.~A. Mohammed, and S.~Al-Zawi, ``Understanding of a convolutional
  neural network,'' in \emph{2017 International Conference on Engineering and
  Technology (ICET)}.\hskip 1em plus 0.5em minus 0.4em\relax Ieee, 2017, pp.
  1--6.

\bibitem{anceau2017nanofocused}
S.~Anceau, P.~Bleuet, J.~Cl{\'e}di{\`e}re, L.~Maingault, J.-l. Rainard, and
  R.~Tucoulou, ``Nanofocused x-ray beam to reprogram secure circuits,'' in
  \emph{International Conference on Cryptographic Hardware and Embedded
  Systems}.\hskip 1em plus 0.5em minus 0.4em\relax Springer, 2017, pp.
  175--188.

\bibitem{bai2021targeted}
J.~Bai, B.~Wu, Y.~Zhang, Y.~Li, Z.~Li, and S.-T. Xia, ``Targeted attack against
  deep neural networks via flipping limited weight bits,'' \emph{arXiv preprint
  arXiv:2102.10496}, 2021.

\bibitem{cryptoeprint:2020:1267}
A.~Baksi, S.~Bhasin, J.~Breier, D.~Jap, and D.~Saha, ``Fault attacks in
  symmetric key cryptosystems,'' Cryptology ePrint Archive, Report 2020/1267,
  2020, \url{https://eprint.iacr.org/2020/1267}.

\bibitem{banik2017gift}
S.~Banik, S.~K. Pandey, T.~Peyrin, Y.~Sasaki, S.~M. Sim, and Y.~Todo, ``Gift: a
  small present,'' in \emph{International Conference on Cryptographic Hardware
  and Embedded Systems}.\hskip 1em plus 0.5em minus 0.4em\relax Springer, 2017,
  pp. 321--345.

\bibitem{barenghi2010countermeasures}
A.~Barenghi, L.~Breveglieri, I.~Koren, G.~Pelosi, and F.~Regazzoni,
  ``Countermeasures against fault attacks on software implemented aes:
  effectiveness and cost,'' in \emph{Proceedings of the 5th Workshop on
  Embedded Systems Security}.\hskip 1em plus 0.5em minus 0.4em\relax ACM, 2010,
  p.~7.

\bibitem{csi_nn}
L.~Batina, S.~Bhasin, D.~Jap, and S.~Picek, ``$\{$CSI$\}$$\{$NN$\}$: Reverse
  engineering of neural network architectures through electromagnetic side
  channel,'' in \emph{28th $\{$USENIX$\}$ Security Symposium ($\{$USENIX$\}$
  Security 19)}, 2019, pp. 515--532.

\bibitem{biggio2013evasion}
B.~Biggio, I.~Corona, D.~Maiorca, B.~Nelson, N.~{\v{S}}rndi{\'c}, P.~Laskov,
  G.~Giacinto, and F.~Roli, ``Evasion attacks against machine learning at test
  time,'' in \emph{Joint European conference on machine learning and knowledge
  discovery in databases}.\hskip 1em plus 0.5em minus 0.4em\relax Springer,
  2013, pp. 387--402.

\bibitem{biggio2012poisoning}
B.~Biggio, B.~Nelson, and P.~Laskov, ``Poisoning attacks against support vector
  machines,'' \emph{arXiv preprint arXiv:1206.6389}, 2012.

\bibitem{biggio2018wild}
B.~Biggio and F.~Roli, ``Wild patterns: Ten years after the rise of adversarial
  machine learning,'' \emph{Pattern Recognition}, vol.~84, pp. 317--331, 2018.

\bibitem{boles2017voice}
A.~Boles and P.~Rad, ``Voice biometrics: Deep learning-based voiceprint
  authentication system,'' in \emph{2017 12th System of Systems Engineering
  Conference (SoSE)}.\hskip 1em plus 0.5em minus 0.4em\relax IEEE, 2017, pp.
  1--6.

\bibitem{bozzato2019shaping}
C.~Bozzato, R.~Focardi, and F.~Palmarini, ``Shaping the glitch: optimizing
  voltage fault injection attacks,'' \emph{IACR Transactions on Cryptographic
  Hardware and Embedded Systems}, pp. 199--224, 2019.

\bibitem{automated_book}
J.~Breier, X.~Hou, and S.~Bhasin, Eds., \emph{Automated Methods in
  Cryptographic Fault Analysis}, 1st~ed.\hskip 1em plus 0.5em minus 0.4em\relax
  Springer, Mar 2019.

\bibitem{breier2018practical}
J.~Breier, X.~Hou, D.~Jap, L.~Ma, S.~Bhasin, and Y.~Liu, ``Practical fault
  attack on deep neural networks,'' in \emph{Proceedings of the 2018 ACM SIGSAC
  Conference on Computer and Communications Security}, 2018, pp. 2204--2206.

\bibitem{breier2015testing}
J.~Breier and D.~Jap, ``Testing feasibility of back-side laser fault injection
  on a microcontroller,'' in \emph{Proceedings of the WESS'15: Workshop on
  Embedded Systems Security}, 2015, pp. 1--6.

\bibitem{breier2015laser}
J.~Breier, D.~Jap, and C.-N. Chen, ``Laser profiling for the back-side fault
  attacks: with a practical laser skip instruction attack on aes,'' in
  \emph{Proceedings of the 1st ACM Workshop on Cyber-Physical System Security},
  2015, pp. 99--103.

\bibitem{breier2020sniff}
J.~Breier, D.~Jap, X.~Hou, S.~Bhasin, and Y.~Liu, ``Sniff: reverse engineering
  of neural networks with fault attacks,'' \emph{arXiv preprint
  arXiv:2002.11021}, 2020.

\bibitem{chen2018detecting}
B.~Chen, W.~Carvalho, N.~Baracaldo, H.~Ludwig, B.~Edwards, T.~Lee, I.~Molloy,
  and B.~Srivastava, ``Detecting backdoor attacks on deep neural networks by
  activation clustering,'' \emph{arXiv preprint arXiv:1811.03728}, 2018.

\bibitem{chen2017targeted}
X.~Chen, C.~Liu, B.~Li, K.~Lu, and D.~Song, ``Targeted backdoor attacks on deep
  learning systems using data poisoning,'' \emph{arXiv preprint
  arXiv:1712.05526}, 2017.

\bibitem{deng2012mnist}
L.~Deng, ``The mnist database of handwritten digit images for machine learning
  research [best of the web],'' \emph{IEEE Signal Processing Magazine},
  vol.~29, no.~6, pp. 141--142, 2012.

\bibitem{deng2020analysis}
Y.~Deng, X.~Zheng, T.~Zhang, C.~Chen, G.~Lou, and M.~Kim, ``An analysis of
  adversarial attacks and defenses on autonomous driving models,'' in
  \emph{2020 IEEE International Conference on Pervasive Computing and
  Communications (PerCom)}.\hskip 1em plus 0.5em minus 0.4em\relax IEEE, 2020,
  pp. 1--10.

\bibitem{gruss2016rowhammer}
D.~Gruss, C.~Maurice, and S.~Mangard, ``Rowhammer. js: A remote
  software-induced fault attack in javascript,'' in \emph{International
  conference on detection of intrusions and malware, and vulnerability
  assessment}.\hskip 1em plus 0.5em minus 0.4em\relax Springer, 2016, pp.
  300--321.

\bibitem{gu2017badnets}
T.~Gu, B.~Dolan-Gavitt, and S.~Garg, ``Badnets: Identifying vulnerabilities in
  the machine learning model supply chain,'' \emph{arXiv preprint
  arXiv:1708.06733}, 2017.

\bibitem{guo2021overview}
W.~Guo, B.~Tondi, and M.~Barni, ``An overview of backdoor attacks against deep
  neural networks and possible defences,'' \emph{arXiv preprint
  arXiv:2111.08429}, 2021.

\bibitem{hecht1992theory}
R.~Hecht-Nielsen, ``Theory of the backpropagation neural network,'' in
  \emph{Neural networks for perception}.\hskip 1em plus 0.5em minus 0.4em\relax
  Elsevier, 1992, pp. 65--93.

\bibitem{hong2019terminal}
S.~Hong, P.~Frigo, Y.~Kaya, C.~Giuffrida, and T.~Dumitraș, ``Terminal brain
  damage: Exposing the graceless degradation in deep neural networks under
  hardware fault attacks,'' in \emph{28th $\{$USENIX$\}$ Security Symposium
  ($\{$USENIX$\}$ Security 19)}, 2019, pp. 497--514.

\bibitem{hou2021physical}
X.~Hou, J.~Breier, D.~Jap, L.~Ma, S.~Bhasin, and Y.~Liu, ``Physical security of
  deep learning on edge devices: Comprehensive evaluation of fault injection
  attack vectors,'' \emph{Microelectronics Reliability}, vol. 120, p. 114116,
  2021.

\bibitem{jagielski2019high}
M.~Jagielski, N.~Carlini, D.~Berthelot, A.~Kurakin, and N.~Papernot,
  ``High-fidelity extraction of neural network models,'' \emph{arXiv preprint
  arXiv:1909.01838}, 2019.

\bibitem{javaid2016deep}
A.~Javaid, Q.~Niyaz, W.~Sun, and M.~Alam, ``A deep learning approach for
  network intrusion detection system,'' in \emph{Proceedings of the 9th EAI
  International Conference on Bio-inspired Information and Communications
  Technologies (formerly BIONETICS)}, 2016, pp. 21--26.

\bibitem{jia2021badencoder}
J.~Jia, Y.~Liu, and N.~Z. Gong, ``Badencoder: Backdoor attacks to pre-trained
  encoders in self-supervised learning,'' \emph{arXiv preprint
  arXiv:2108.00352}, 2021.

\bibitem{joye2012fault}
M.~Joye and M.~Tunstall, \emph{Fault analysis in cryptography}.\hskip 1em plus
  0.5em minus 0.4em\relax Springer, 2012, vol. 147.

\bibitem{kenjar2020v0ltpwn}
Z.~Kenjar, T.~Frassetto, D.~Gens, M.~Franz, and A.-R. Sadeghi, ``V0ltpwn:
  Attacking x86 processor integrity from software,'' in \emph{29th
  $\{$USENIX$\}$ Security Symposium ($\{$USENIX$\}$ Security 20)}, 2020, pp.
  1445--1461.

\bibitem{kolosnjaji2016deep}
B.~Kolosnjaji, A.~Zarras, G.~Webster, and C.~Eckert, ``Deep learning for
  classification of malware system call sequences,'' in \emph{Australasian
  Joint Conference on Artificial Intelligence}.\hskip 1em plus 0.5em minus
  0.4em\relax Springer, 2016, pp. 137--149.

\bibitem{krautter2018fpgahammer}
J.~Krautter, D.~R. Gnad, and M.~B. Tahoori, ``Fpgahammer: Remote voltage fault
  attacks on shared fpgas, suitable for dfa on aes,'' \emph{IACR Transactions
  on Cryptographic Hardware and Embedded Systems}, pp. 44--68, 2018.

\bibitem{krizhevsky2012imagenet}
A.~Krizhevsky, I.~Sutskever, and G.~E. Hinton, ``Imagenet classification with
  deep convolutional neural networks,'' \emph{Advances in neural information
  processing systems}, vol.~25, pp. 1097--1105, 2012.

\bibitem{lecun1998gradient}
Y.~LeCun, L.~Bottou, Y.~Bengio, and P.~Haffner, ``Gradient-based learning
  applied to document recognition,'' \emph{Proceedings of the IEEE}, vol.~86,
  no.~11, pp. 2278--2324, 1998.

\bibitem{li2018learning}
H.~Li, K.~Ota, and M.~Dong, ``Learning iot in edge: Deep learning for the
  internet of things with edge computing,'' \emph{IEEE network}, vol.~32,
  no.~1, pp. 96--101, 2018.

\bibitem{li2020backdoor}
Y.~Li, B.~Wu, Y.~Jiang, Z.~Li, and S.-T. Xia, ``Backdoor learning: A survey,''
  \emph{arXiv preprint arXiv:2007.08745}, 2020.

\bibitem{li2021survey}
Y.~Li, H.~Wang, and M.~Barni, ``A survey of deep neural network watermarking
  techniques,'' \emph{arXiv preprint arXiv:2103.09274}, 2021.

\bibitem{liao2018backdoor}
C.~Liao, H.~Zhong, A.~Squicciarini, S.~Zhu, and D.~Miller, ``Backdoor embedding
  in convolutional neural network models via invisible perturbation,''
  \emph{arXiv preprint arXiv:1808.10307}, 2018.

\bibitem{liu2018fine}
K.~Liu, B.~Dolan-Gavitt, and S.~Garg, ``Fine-pruning: Defending against
  backdooring attacks on deep neural networks,'' in \emph{International
  Symposium on Research in Attacks, Intrusions, and Defenses}.\hskip 1em plus
  0.5em minus 0.4em\relax Springer, 2018, pp. 273--294.

\bibitem{liu2017fault}
Y.~Liu, L.~Wei, B.~Luo, and Q.~Xu, ``Fault injection attack on deep neural
  network,'' in \emph{Proceedings of the 36th International Conference on
  Computer-Aided Design}.\hskip 1em plus 0.5em minus 0.4em\relax IEEE Press,
  2017, pp. 131--138.

\bibitem{liu2019abs}
Y.~Liu, W.-C. Lee, G.~Tao, S.~Ma, Y.~Aafer, and X.~Zhang, ``Abs: Scanning
  neural networks for back-doors by artificial brain stimulation,'' in
  \emph{Proceedings of the 2019 ACM SIGSAC Conference on Computer and
  Communications Security}, 2019, pp. 1265--1282.

\bibitem{liu2017trojaning}
Y.~Liu, S.~Ma, Y.~Aafer, W.-C. Lee, J.~Zhai, W.~Wang, and X.~Zhang, ``Trojaning
  attack on neural networks,'' 2017.

\bibitem{moro2014formal}
N.~Moro, K.~Heydemann, E.~Encrenaz, and B.~Robisson, ``Formal verification of a
  software countermeasure against instruction skip attacks,'' \emph{Journal of
  Cryptographic Engineering}, vol.~4, no.~3, pp. 145--156, 2014.

\bibitem{munoz2017towards}
L.~Mu{\~n}oz-Gonz{\'a}lez, B.~Biggio, A.~Demontis, A.~Paudice, V.~Wongrassamee,
  E.~C. Lupu, and F.~Roli, ``Towards poisoning of deep learning algorithms with
  back-gradient optimization,'' in \emph{Proceedings of the 10th ACM Workshop
  on Artificial Intelligence and Security}, 2017, pp. 27--38.

\bibitem{murdock2020plundervolt}
K.~Murdock, D.~Oswald, F.~D. Garcia, J.~Van~Bulck, F.~Piessens, and D.~Gruss,
  ``Plundervolt: How a little bit of undervolting can create a lot of
  trouble,'' \emph{IEEE Security \& Privacy}, vol.~18, no.~5, pp. 28--37, 2020.

\bibitem{narodytska2017simple}
N.~Narodytska and S.~P. Kasiviswanathan, ``Simple black-box adversarial attacks
  on deep neural networks.'' in \emph{CVPR Workshops}, vol.~2, 2017.

\bibitem{papadimitriou1981complexity}
C.~H. Papadimitriou, ``On the complexity of integer programming,''
  \emph{Journal of the ACM (JACM)}, vol.~28, no.~4, pp. 765--768, 1981.

\bibitem{papernot2016practical}
N.~Papernot, P.~McDaniel, I.~Goodfellow, S.~Jha, Z.~B. Celik, and A.~Swami,
  ``Practical black-box attacks against deep learning systems using adversarial
  examples,'' \emph{arXiv preprint arXiv:1602.02697}, vol.~1, no.~2, p.~3,
  2016.

\bibitem{patrick2016lightweight}
C.~Patrick, B.~Yuce, N.~F. Ghalaty, and P.~Schaumont, ``Lightweight fault
  attack resistance in software using intra-instruction redundancy,'' in
  \emph{International Conference on Selected Areas in Cryptography}.\hskip 1em
  plus 0.5em minus 0.4em\relax Springer, 2016, pp. 231--244.

\bibitem{qiu2019voltjockey}
P.~Qiu, D.~Wang, Y.~Lyu, and G.~Qu, ``Voltjockey: Breaching trustzone by
  software-controlled voltage manipulation over multi-core frequencies,'' in
  \emph{Proceedings of the 2019 ACM SIGSAC Conference on Computer and
  Communications Security}, 2019, pp. 195--209.

\bibitem{rakin2020tbt}
A.~S. Rakin, Z.~He, and D.~Fan, ``Tbt: Targeted neural network attack with bit
  trojan,'' in \emph{Proceedings of the IEEE/CVF Conference on Computer Vision
  and Pattern Recognition}, 2020, pp. 13\,198--13\,207.

\bibitem{rehman2019backdoor}
H.~Rehman, A.~Ekelhart, and R.~Mayer, ``Backdoor attacks in neural networks--a
  systematic evaluation on multiple traffic sign datasets,'' in
  \emph{International Cross-Domain Conference for Machine Learning and
  Knowledge Extraction}.\hskip 1em plus 0.5em minus 0.4em\relax Springer, 2019,
  pp. 285--300.

\bibitem{simonyan2014very}
K.~Simonyan and A.~Zisserman, ``Very deep convolutional networks for
  large-scale image recognition,'' \emph{arXiv preprint arXiv:1409.1556}, 2014.

\bibitem{szegedy2013intriguing}
C.~Szegedy, W.~Zaremba, I.~Sutskever, J.~Bruna, D.~Erhan, I.~Goodfellow, and
  R.~Fergus, ``Intriguing properties of neural networks,'' \emph{arXiv preprint
  arXiv:1312.6199}, 2013.

\bibitem{szegedy2014intriguing}
------, ``Intriguing properties of neural networks,'' in \emph{ICLR}, 2014.

\bibitem{tunstall2011differential}
M.~Tunstall, D.~Mukhopadhyay, and S.~Ali, ``Differential fault analysis of the
  advanced encryption standard using a single fault,'' in \emph{IFIP
  international workshop on information security theory and practices}.\hskip
  1em plus 0.5em minus 0.4em\relax Springer, 2011, pp. 224--233.

\bibitem{wang2019neural}
B.~Wang, Y.~Yao, S.~Shan, H.~Li, B.~Viswanath, H.~Zheng, and B.~Y. Zhao,
  ``Neural cleanse: Identifying and mitigating backdoor attacks in neural
  networks,'' in \emph{2019 IEEE Symposium on Security and Privacy (SP)}.\hskip
  1em plus 0.5em minus 0.4em\relax IEEE, 2019, pp. 707--723.

\bibitem{wang2018great}
B.~Wang, Y.~Yao, B.~Viswanath, H.~Zheng, and B.~Y. Zhao, ``With great training
  comes great vulnerability: Practical attacks against transfer learning,'' in
  \emph{27th $\{$USENIX$\}$ Security Symposium ($\{$USENIX$\}$ Security 18)},
  2018, pp. 1281--1297.

\bibitem{wang2018data}
Y.~Wang and K.~Chaudhuri, ``Data poisoning attacks against online learning,''
  \emph{arXiv preprint arXiv:1808.08994}, 2018.

\bibitem{xiao2015feature}
H.~Xiao, B.~Biggio, G.~Brown, G.~Fumera, C.~Eckert, and F.~Roli, ``Is feature
  selection secure against training data poisoning?'' in \emph{International
  Conference on Machine Learning}.\hskip 1em plus 0.5em minus 0.4em\relax PMLR,
  2015, pp. 1689--1698.

\bibitem{xu2021security}
Q.~Xu, M.~T. Arafin, and G.~Qu, ``Security of neural networks from hardware
  perspective: A survey and beyond,'' in \emph{2021 26th Asia and South Pacific
  Design Automation Conference (ASP-DAC)}.\hskip 1em plus 0.5em minus
  0.4em\relax IEEE, 2021, pp. 449--454.

\bibitem{zhang2021backdoor}
Z.~Zhang, J.~Jia, B.~Wang, and N.~Z. Gong, ``Backdoor attacks to graph neural
  networks,'' in \emph{Proceedings of the 26th ACM Symposium on Access Control
  Models and Technologies}, 2021, pp. 15--26.

\bibitem{zhao2019fault}
P.~Zhao, S.~Wang, C.~Gongye, Y.~Wang, Y.~Fei, and X.~Lin, ``Fault sneaking
  attack: A stealthy framework for misleading deep neural networks,'' in
  \emph{2019 56th ACM/IEEE Design Automation Conference (DAC)}.\hskip 1em plus
  0.5em minus 0.4em\relax IEEE, 2019, pp. 1--6.

\end{thebibliography}

\end{document}